# Deciphering the Nanoscale Frictional Properties of Two-Dimensional SnSe and $SnSe_2$ with Lateral Force Microscopy

M. Ozdogan[1, ‡], T. Iken[1, ‡], D. Cakir[1,*], N. Oncel[1,*]
*Department of Physics and Astrophysics, University of North Dakota, Grand Forks, ND 58202, United States of America*

*Corresponding authors: deniz.cakir@und.edu ; nuri.oncel@und.edu
[‡]Equal contribution

**Abstract:** In this study, we employed lateral force microscopy, a specialized technique within atomic force microscopy, to explore the nanoscale frictional properties of 2D SnSe and $SnSe_2$ layers, and we compared these findings with graphene, a commonly used solid lubricant. Our results revealed that SnSe and $SnSe_2$ layers exhibited superior lubricative performance compared to graphene with similar thickness, with $SnSe_2$ demonstrating friction levels up to three times lower. This lower friction observed in $SnSe_2$, as compared to SnSe, was further substantiated by extensive Density Functional Theory (DFT) calculations, confirming the intrinsic material properties contributing to their enhanced lubricative behavior.

## Introduction

Understanding and controlling nanoscale friction holds promise for reducing energy consumption and boosting efficiency across diverse engineering applications.[1] Liquid lubricants effectively reduce friction in many conventional mechanical systems, but they often fall short in nano- and micro-mechanical systems due to excessive viscosity or environmental contamination.[2] In such scenarios, solid lubricants, adept at operating under extreme conditions, emerge as a more favorable solution, providing reliability and superior performance.[3] Furthermore, layered 2D materials are particularly effective in the regime of solid lubricants because they have weaker van der Waals forces (vdW) between adjacent atomic layers that facilitate easy sliding.[4] Transition metal sulfides and selenides such as $MoS_2$ have very low friction coefficients in dry and vacuum environments, making them suitable candidates for space explorations.[2] Graphite has been used since ancient times due to its effectiveness in humid environments.[5] Therefore, graphene and $MoS_2$ have been extensively studied as solid lubricants, but the frictional properties of other 2D materials deserve to be explored.

Among those 2D materials are SnSe and $SnSe_2$. Recently, SnSe has garnered significant attention due to its exceptional thermoelectric performance.[6] It comprises a layered orthorhombic crystal structure with 0.6 nm thick layers (see Figure 1(a)).[7] Characterized by strong in-plane bonds and, conversely, weaker vdW forces between the layers facilitate SnSe for easy cleavage[8] and contribute to its solid lubricating capabilities, a property closely tied to its crystallographic arrangement.[9] Similarly, $SnSe_2$, emerging as a material of significant interest for applications in solar cells and data storage,[10] is characterized by a hexagonal layered structure with layer thicknesses of approximately 0.65 nm (see Figure 1(b)). It also features strong in-plane covalent bonds and relatively weak interlayer vdW interactions, which ease layer separation and cleavage. Its tribological properties, particularly its low shear force relative to other 2D materials, make $SnSe_2$ a promising candidate for low friction at the nanoscale, positioning it as an attractive material for superlubricity research.[10]

Building on this motivation and, to our knowledge, as the first to explore this area, we employed lateral force microscopy (LFM), a specialized atomic force microscopy (AFM) technique, to study the nanoscale frictional properties of 2D SnSe and $SnSe_2$. We then compared these findings with graphene flakes and bulk graphite, a commonly used solid lubricant.

Additionally, we conducted ab initio density functional theory calculations to establish a correlation between the experimental results and the materials' intrinsic properties.

## Experimental

$SnSe_2$ samples were grown inside a 1-inch quartz tube using a horizontal furnace. Sn and Se powders ( $\sim 100\, mesh, \geq$ 99.5%, Sigma Aldrich) were mixed in a stoichiometric ratio and placed at the center of the quartz tube via an alumina boat, while freshly cleaved mica substrates were placed downstream at $10 - 15\ cm$ away from the center. The quartz tube was pumped down to a base pressure of $10^{-6}\ mbar$ and refilled three times with ultra-pure Ar gas at a pressure slightly above the atmospheric conditions to minimize oxygen contamination. The furnace was heated to $500\text{–}550\ °C$ for 50 min, kept at this temperature for 10 min with a growth pressure of 100 mbar in Ar flow, then left to cool down to room temperature. The composition of $SnSe_2$ was confirmed using energy-dispersive X-ray (EDX) spectroscopy. On the other hand, SnSe layers were mechanically exfoliated from bulk pieces (SnSe, 99.999%, ThermoFisher Scientific) and transferred onto the substrates via scotch tape. For both samples, $300\ nm$ thick thermally grown $SiO_2$ substrates were used.

To investigate the nanoscale friction properties of both 2D SnSe and $SnSe_2$ layers, we performed LFM measurements using Bruker's NanoWizard-4XP series AFM. All measurements were conducted under atmospheric conditions at room temperature in an acoustically and vibrationally isolated chamber to minimize external interferences. LFM measurements were performed using Bruker's ESPA-V2 probe with a tip radius of $\sim 8\ nm$, a nominal resonance frequency of $13\ kHz$, and a spring constant of $0.2\ N/m$. Each cantilever's normal and lateral calibrations were done separately by thermal noise and improved wedge methods,[11] respectively.

The latter was performed by scanning the commercial MikroMasch TGF 11 silicon calibration sample (NanoAndMore Corp.). The lateral force constants were found between $32\ to\ 50\ nN/V$. The scanning frequency was kept at 1 Hz during all measurements, and the normal load applied was $1.0\ nN$. Measurements were performed in contact mode, and topography maps were recorded simultaneously with lateral force maps, from which friction maps were generated. Corresponding topography and friction maps were post-processed using Gwyddion, an open-source software.[12] Each image was processed using mean filtering to level the plane and align the rows, removing the background.

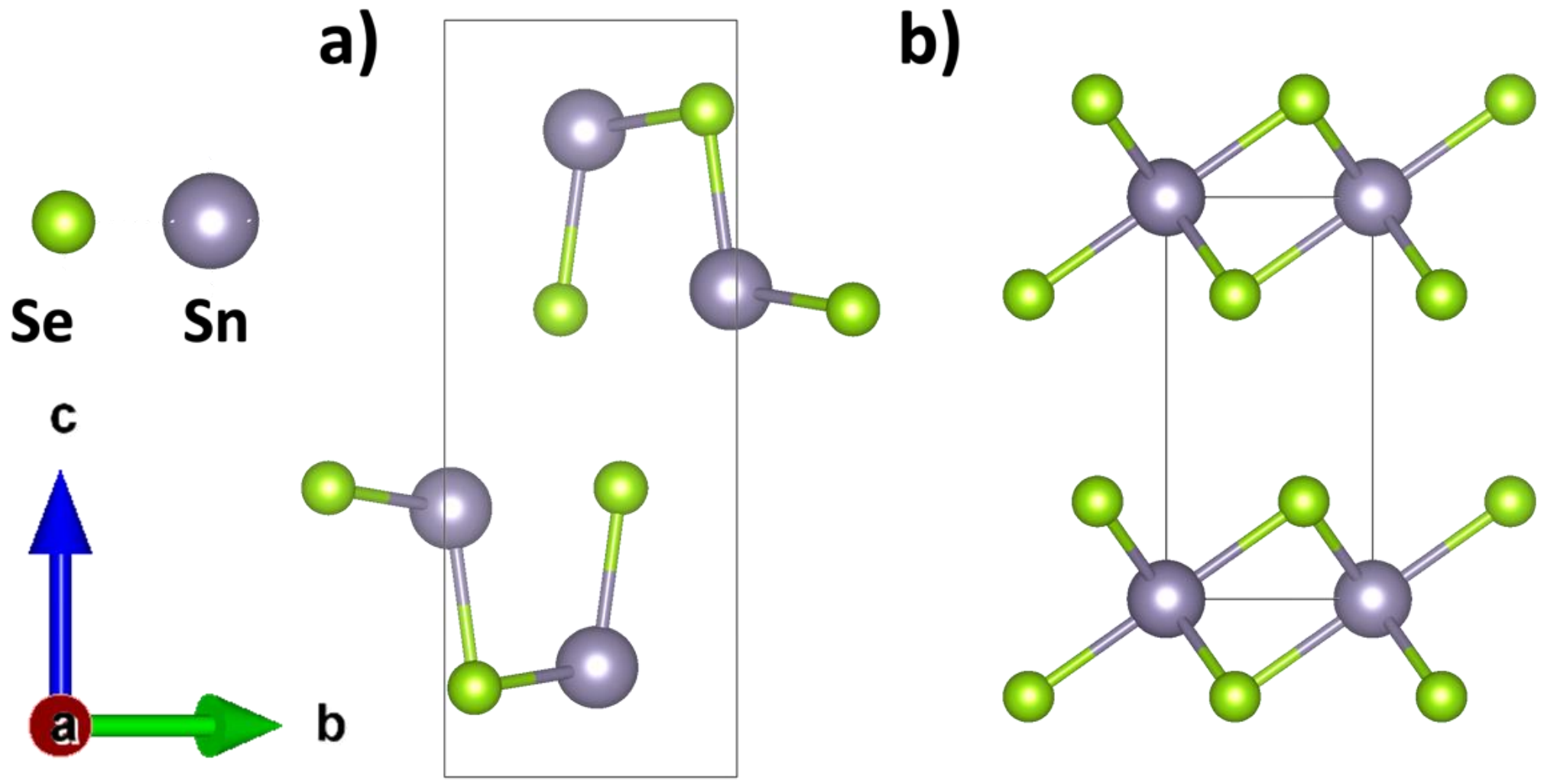

**Figure 1** Crystal structure of **(a)** SnSe and **(b)** $SnSe_2$. Black rectangles represent single unit cells.

## Computational

We conducted a thorough computational study on SnSe and $SnSe_2$, examining monolayer, bilayer, and bulk systems. The Vienna Ab-initio Simulation Package (VASP) was used for all calculations,[13] utilizing the Generalized Gradient Approximation (GGA) with the Perdew, Burke, and Ernzerhof (PBE) formalism[14] for the exchange-correlation functional and the Projector Augmented-Wave (PAW) method for the potentials.[15] An energy cutoff of $500\ eV$ was employed throughout. To accurately account for vdW interactions, crucial for layered materials, we tested three different vdW correction schemes: the empirical corrections DFT+D3,[16] DFT+D2,[17] and optB86b-vdW functional.[18] We compared the calculated lattice parameters obtained from each functional with experimental values to determine the most appropriate vdW correction for our systems.

The calculations involved structural optimization, allowing all atomic positions and lattice parameters to relax until the forces on all atoms were less than $0.01\ eV/\text{Å}$ and the total energy change was less than $10^{-6} eV$. For the $\boldsymbol{k}$-point sampling of the Brillouin zone, a $16 \times 16 \times 1$ $\boldsymbol{k}$-mesh was used for monolayer and bilayer systems, while a $16 \times 16 \times 6$ $\boldsymbol{k}$-mesh was applied for bulk systems to ensure sufficient sampling for obtaining accurate lattice parameters. We compared the computed lattice parameters with experimental data to identify the best vdW correction to describe the interlayer interactions in SnSe and $SnSe_2$. This benchmarking is essential to ensure the accuracy of further computational studies on these materials.

## Results & Discussion

The EDX spectroscopy measurements and average atomic percentages were taken at multiple spots on $SnSe_2$ nanoflakes, summarized in Table 1. They confirm the 1:2 ratio of $SnSe_2$ grown by the CVD method. The Si and O signals come from the $SiO_2$ substrate.

**Table 1.** Energy-dispersive X-ray Spectroscopy (EDX) result of $SnSe_2$ nanoflakes, confirming their stoichiometric compositions.

| | **Sn** | **Se** | **O** | **Si** |
|---|---|---|---|---|
| **$SnSe_2$** | 4.37% | 8.19% | 4.86% | 82.58% |

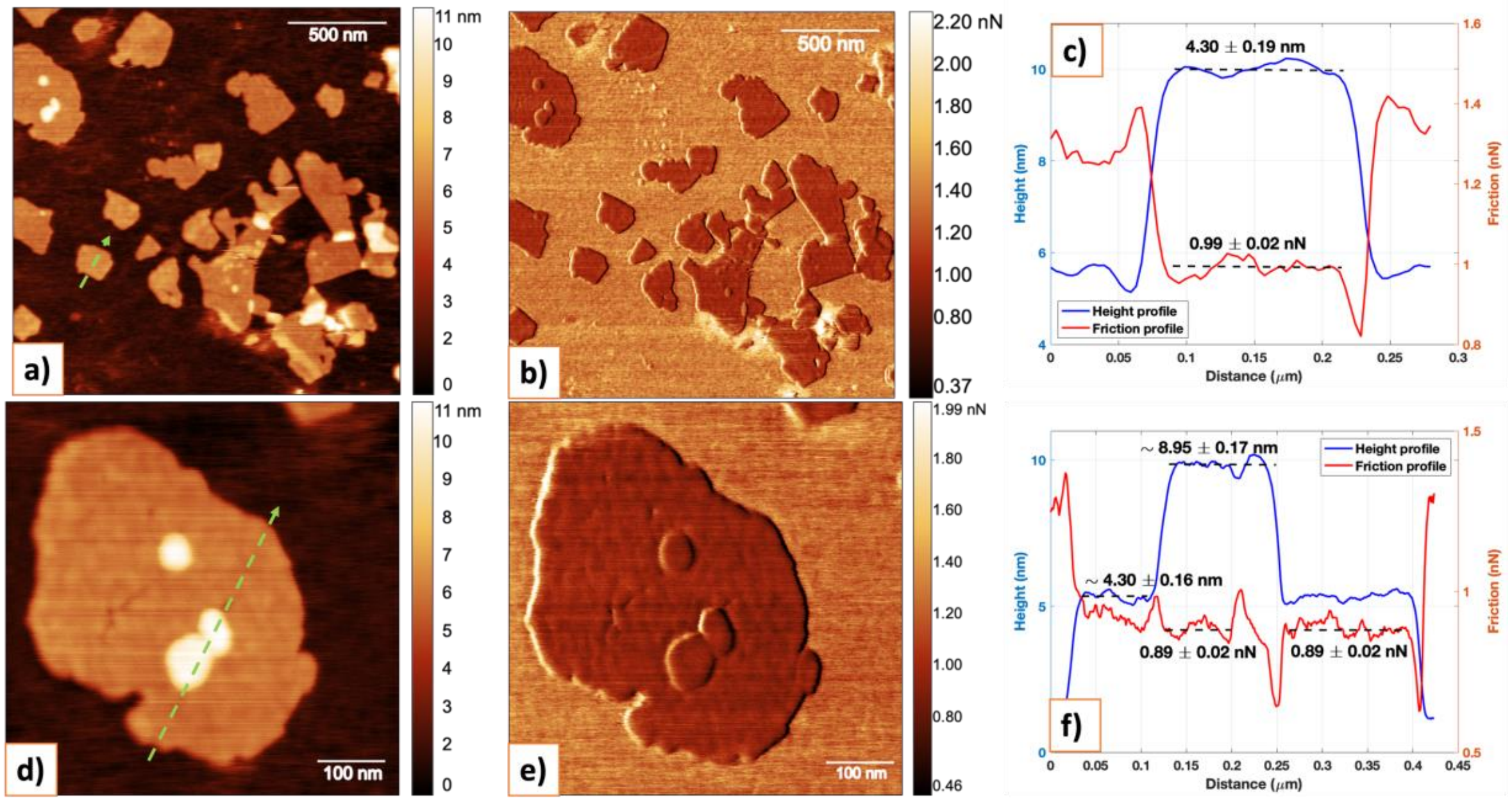

***Figure 2.*** *Summary of nanoscale friction properties of exfoliated SnSe.* ***(a)*** *Topography map of multiple SnSe nanoflakes with similar heights.* ***(b)*** *Corresponding friction map of the same SnSe nanoflakes.* ***(c)*** *Representative height and friction profiles taken from the region were shown by the green dashed arrow in (a), with the arrow showing the profiling direction.* ***(d)*** *Zoomed-in topography map of a representative SnSe nanoflake exhibiting varying heights.* ***(e)*** *Corresponding friction map of the nanoflake in (d).* ***(f)*** *Representative height and friction profiles taken from the region shown by the green dashed arrow in (d), with the arrow indicating the profiling direction.*

To investigate the topographical and frictional characteristics of 2D SnSe and $SnSe_2$ layers, we performed contact mode AFM measurements by recording the probe's vertical and lateral deflections. First, we investigated the exfoliated SnSe layers; Figure 2 summarizes these results. Figure 2(a) shows representative topography maps indicating several SnSe layers with varying lateral sizes but uniform thicknesses of $\sim 4.0\, nm$, corresponding to ~6 atomic layers. By examining the friction map shown in Figure 2(b), we can conclude that SnSe layers with similar thickness exhibit a uniform friction force. To quantify the observed trend, we extracted line profiles (the width of the line profiles was increased to at least $\sim 150\ nm$ to improve data accuracy and better average) across multiple nanoflakes with a similar thickness of $\sim 4.0\ nm$ from at least three different samples and found that, on average, friction force values were $0.98 \pm 0.10\ nN$ (n=33, n is the total number of data points) for exfoliated SnSe samples on $SiO_2$ substrates. Figure 2 (c) compares the typical height and friction profiles taken from a line scan of the same region shown by a green dashed arrow in Figure 2(a). The height of SnSe was approximately 4.30 nm (around 7 layers), with a corresponding average friction force $0.99 \pm 0.02\ nN$ reflecting the overall frictional characteristics of the SnSe layers. Lee *et al.* claimed that 2D materials show decreasing friction with increasing layers.[19] We conducted additional experiments on SnSe flakes with different heights to examine the effect of thickness on friction. Figures 2(d) and 2(e) display the topography and friction maps of a representative SnSe flake with varying thicknesses. As shown in Figure 2(f), the average friction values remained constant even when the thickness was doubled, as indicated by the line profile comparison.

We then shifted our focus to CVD-grown $SnSe_2$ layers and carried out LFM measurements to analyze their frictional properties, as summarized in Figure 3. The topography maps of $SnSe_2$ samples shown in Figure 3(a) exhibit pyramid shapes. It is essential to remember that the topography can cause artifacts in friction measurements, particularly at the edges of flakes. This can result in sudden jumps or dips depending on the scanning direction, as illustrated in the line profile comparisons in Figures 2 and 3. Once the cantilever hits the edges, more lateral twisting leads to unrealistic friction values. To avoid this problem, we measured friction on truncated pyramids highlighted by green dashed circles on the friction map (Figure 3(b)), carefully excluding the edge effects. The average friction force was $0.72 \pm 0.14\ nN$ (n=42) for $SnSe_2$ nanoflakes with an average thickness of 8-10 nm (~10-16 atomic layers). Some of these flakes showed friction forces that were as low as ~0.52 $nN$. Figure 3(e) illustrates an example of these layers with low friction, and their respective topography maps are shown in Figure 3(d). The friction and height profiles drawn over this layer were compared in Figure 3(f). Again, we checked the thickness-dependent friction forces of $SnSe_2$ ranging from 10 to 16 atomic layers, as shown in Figure 3(c). Considering the experimental errors, the friction forces did not show notable variations when the number of $SnSe_2$ layers was changed from 10 to 16.

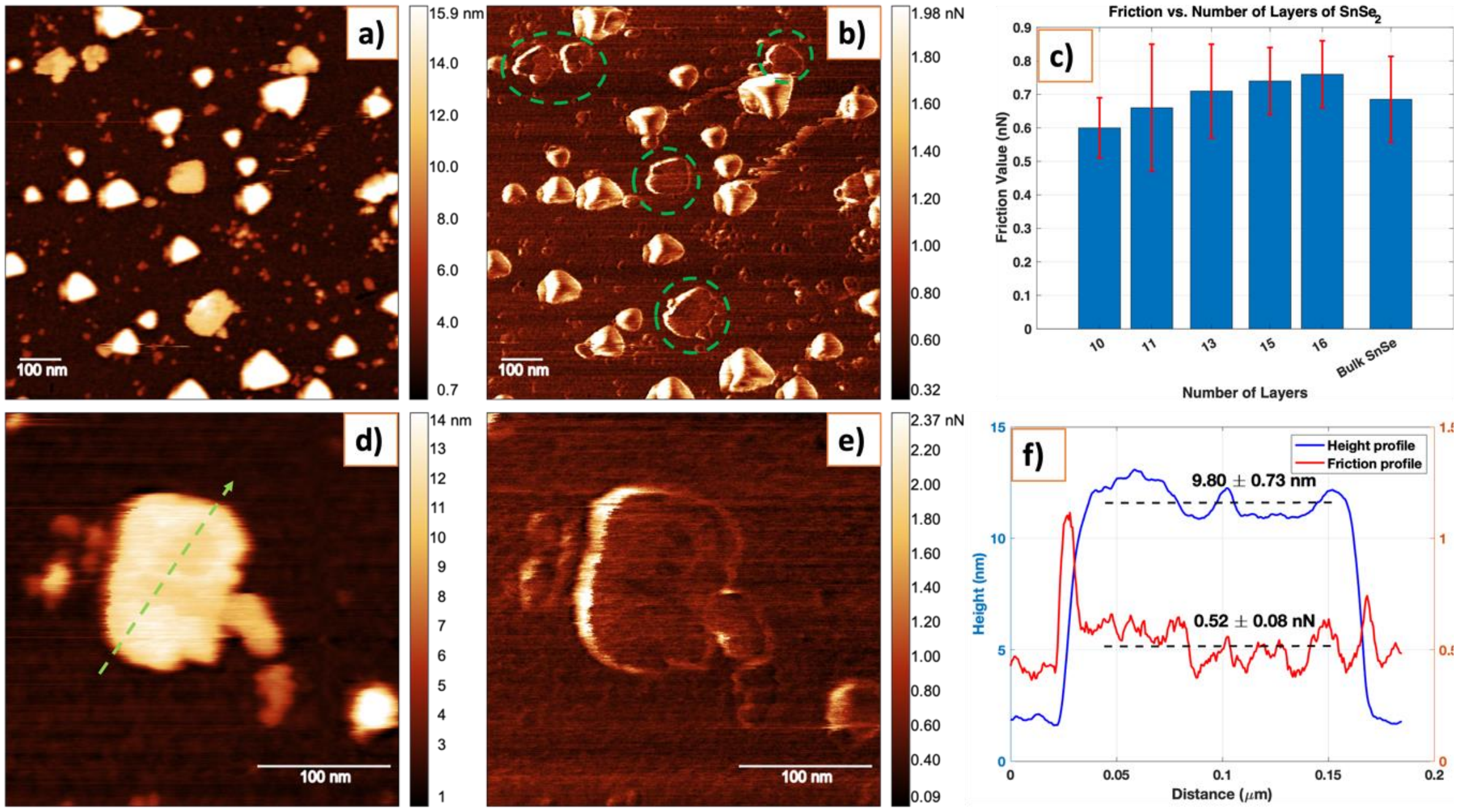

**Figure 3.** Summary of nanoscale friction properties of CVD-grown $SnSe_2$. **(a)** Topography map of multiple $SnSe_2$ nanoflakes. **(b)** Corresponding friction map of the same $SnSe_2$ nanoflakes, green dashed circles indicate the flat flakes where average friction forces were measured. **(c)** Bar plot illustrating the relationship between the number of layers and friction values for $SnSe_2$ nanoflakes, here bulk SnSe friction value also included for comparison **(d)** Closer topography map of a $SnSe_2$ nanoflake showing a low friction value with a thickness of ~10 nm. **(e)** Corresponding friction map of the nanoflake in (d). **(f)** The height and friction profiles taken from the region were shown by the green dashed arrow in (d), with the arrow indicating the profiling direction.

The friction values obtained from LFM experiments can be ambiguous at the nanoscale due to the structure and chemistry of the AFM tips.[20] Therefore, comparing these values with previous works may not be accurate. To provide more reliable results for comparison, we also conducted similar measurements on graphene layers, a widely used solid lubricant material, having similar thicknesses with SnSe and $SnSe_2$ layers. We investigated both mechanically exfoliated graphene layers transferred to a 300 nm thick $SiO_2$ substrate and bulk graphite (HOPG grade ZYB, Nanoscience Instruments). Their average friction forces were found as $1.55 \pm 0.09\, nN$ (n=6) and $0.23 \pm 0.02\, nN$ (n=6), respectively (see supplementary Figure S1). The lubricative performance of SnSe and $SnSe_2$ flakes was superior to that of graphene flakes of similar thickness. Specifically, $SnSe_2$ flakes exhibited nearly three times smaller friction than graphene. However, the bulk friction of SnSe was measured at $0.69 \pm 0.13\, nN$ (n=3) is higher than that of bulk graphite.

To gain a deeper insight into the frictional characteristics of SnSe and $SnSe_2$, we performed a comprehensive computational study on SnSe and $SnSe_2$, considering monolayer, bilayer, and bulk systems. For bulk SnSe (Table 2), the lattice parameters obtained using various vdW functionals indicate that DFT-D3 closely approximates the experimental values, especially for the $c$ lattice parameter. This suggests that DFT-D3 may be the most reliable functional for simulating vdW interactions in this material. DFT-D2 and optB86b slightly overestimate the $a$ and $b$ lattice parameters in the bulk case. In the bilayer setup, DFT-D2 predicts the strongest interlayer binding energy at $40.2\, meV/Å^2$, whereas DFT-D3 and optB86b offer slightly lower values at $33.4$ and $34.8\, meV/Å^2$, respectively. In monolayer SnSe, different functionals show significant variation in the $b$ lattice parameter, contrary to the bilayer and bulk case, with DFT-D3 predicting a notably larger value.

**Table 2** Lattice parameters for bulk, bilayer, and monolayer SnSe were computed using different vdW corrections, and the experimental parameters were used for comparison. Furthermore, interlayer binding energies were calculated for the bilayer system.

| SnSe | | ***a***(Å) | ***b***(Å) | ***c***(Å) | $E_{bind}$ (meV/Å$^2$) |
|---|---|---|---|---|---|
| **Bulk** | DFT-D2 | 4.186 | 4.401 | 11.628 | |
| | DFT-D3 | 4.172 | 4.377 | 11.488 | |
| | optB86b | 4.196 | 4.426 | 11.625 | |
| | Experiment[21] | 4.135 | 4.440 | 11.490 | |
| | Experiment[22] | 4.153 | 4.450 | 11.502 | |
| **Bilayer** | DFT-D2 | 4.216 | 4.375 | | 40.2 |
| | DFT-D3 | 4.213 | 4.475 | | 33.4 |
| | optB86b | 4.233 | 4.370 | | 34.8 |
| **Monolayer** | DFT-D2 | 4.244 | 4.340 | | |
| | DFT-D3 | 4.268 | 4.505 | | |
| | optB86b | 4.262 | 4.326 | | |

In bulk $SnSe_2$ (Table 3), DFT-D2 provides lattice parameters closest to the experimental values, making it the most reliable for bulk calculations. While all three functionals slightly overestimate the $a$ parameter, DFT-D2, and optB86b offer the best approximation for the $c$ parameter. In the bilayer system, the most reliable interlayer binding energy was predicted to be $22.2\ meV/Å^2$ by DFT-D3, closely followed by optB86b with $21.3\ meV/Å^2$, and DFT-D2 with $18.4\ meV/Å^2$. It is worth noting that for $SnSe_2$ compared to SnSe, DFT-D3 might slightly overestimate van der Waals interactions. However, regardless of the vdW correction used when the binding energies between different layers of SnSe and $SnSe_2$ are compared, SnSe exhibits the strongest interlayer interaction.

**Table 2.** Lattice parameters for bulk, bilayer, and monolayer $SnSe_2$ were computed using different vdW corrections, and the experimental parameters were used for comparison. Furthermore, interlayer binding energies were calculated for the bilayer system.

| $SnSe_2$ | | $a$(Å) | $c$(Å) | $E_{bind}$ (meV/Å$^2$) |
|---|---|---|---|---|
| **Bulk** | DFT-D2 | 3.834 | 6.156 | |
| | DFT-D3 | 3.841 | 6.006 | |
| | optB86b | 3.858 | 6.108 | |
| | Experiment[21] | 3.810 | 6.140 | |
| **Bilayer** | DFT-D2 | 3.829 | | 18.4 |
| | DFT-D3 | 3.835 | | 22.2 |
| | optB86b | 3.848 | | 21.3 |
| **Monolayer** | DFT-D2 | 3.824 | | |
| | DFT-D3 | 3.825 | | |
| | optB86b | 3.834 | | |

To comprehensively understand the frictional behavior, we constructed the potential energy surface (PES) by calculating the interlayer interaction energy for bilayer SnSe and $SnSe_2$, as depicted in Figure 4. The PES visualizes the atomic surface roughness of the 2D materials, where each atom's local force field forms energy barriers separated by interatomic energy minima. Three different scenarios were considered for these calculations: (i) the top layer slides over the bottom layer at the equilibrium interlayer separation without any relaxation, (ii) the topmost and bottommost atoms are fixed, allowing all other atoms to relax freely, (iii) relaxation is permitted only in the out-of-plane (z-direction). Figure 4(a) illustrates the setup of these three scenarios, while Figures 4(b) and 4(c) present the resulting PES for bilayer SnSe and $SnSe_2$, respectively. The black arrows indicate the sliding direction of the top layer, and the white lines represent the minimum energy paths for sliding. Across all scenarios, the PES patterns are similar for both SnSe and $SnSe_2$ bilayers. However, the results show that the friction between SnSe layers is significantly higher than that of $SnSe_2$ layers, aligning with the calculated adhesion energies. This result is also consistent with our experimental results, which confirm that $SnSe_2$ has a lower friction coefficient than SnSe.

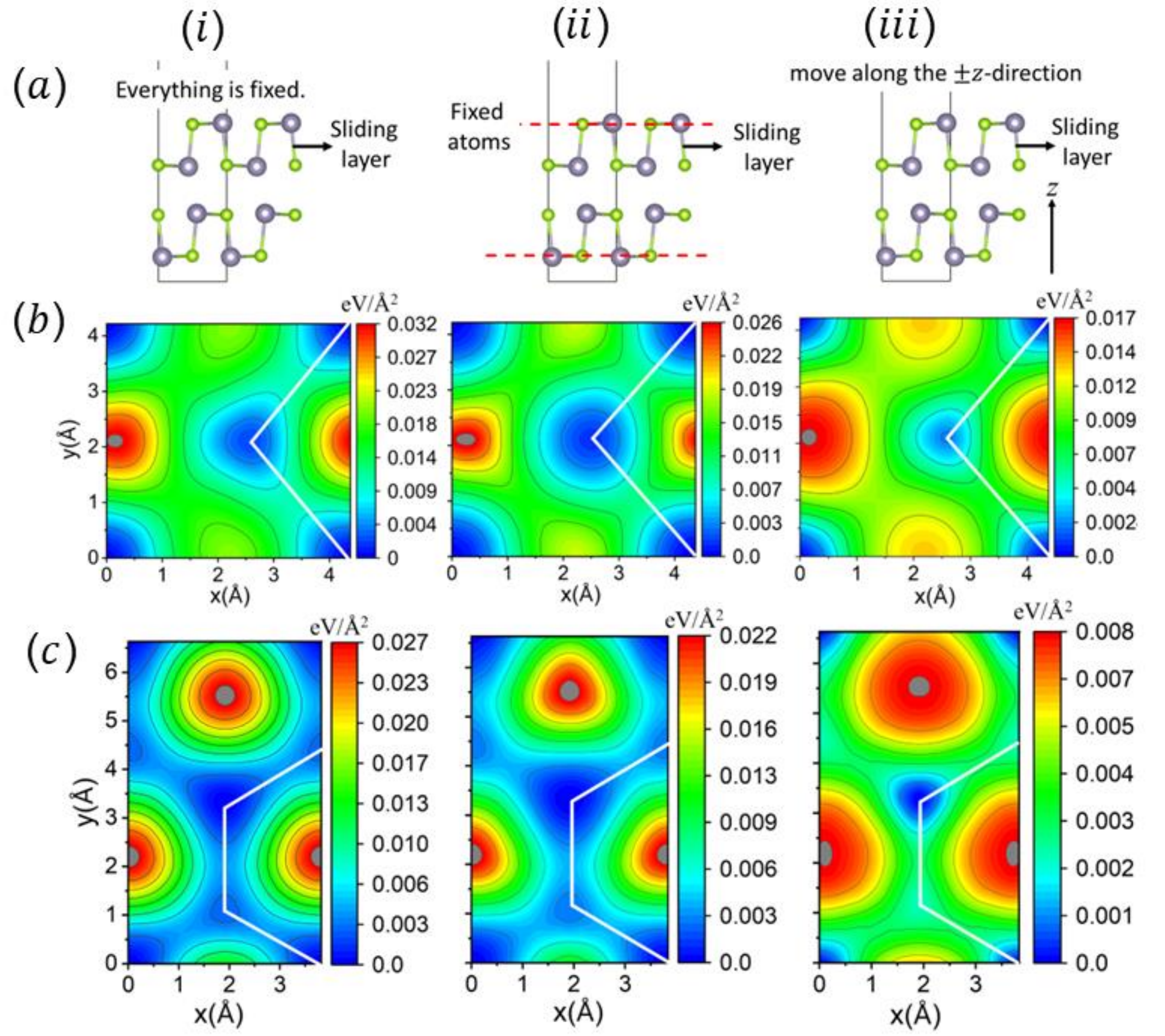

**Figure 4**. Potential energy surface (PES) analysis for bilayer SnSe and $SnSe_2$ under three scenarios. **(a)** Schematic representation of the three scenarios used in the PES calculations: (i) sliding of the top layer over the bottom layer at the equilibrium interlayer separation without relaxation, (ii) sliding with fixed topmost and bottommost atoms, allowing other atoms to relax, and (iii) sliding with relaxation allowed only in the out-of-plane (z-direction). **(b)** PES plots for bilayer SnSe corresponding to scenarios (i), (ii), and (iii), respectively. **(c)** PES plots for bilayer $SnSe_2$ corresponding to scenarios (i), (ii), and (iii), respectively. The black arrows indicate the sliding direction of the top layer, and the white lines represent the minimum energy paths for the sliding process.

To simulate the friction process observed in our AFM experiments, we employed a single methane ($CH_4$) molecule as a probe. To avoid interactions between neighboring $CH_4$ molecules, we used a rectangular supercell model (3 × 3) for SnSe and $SnSe_2$ monolayers and a vacuum region of 15 Å in the z direction to eliminate interactions between adjacent supercells. The sliding energy was calculated for two different separations between the $CH_4$ molecule and the surfaces of SnSe and $SnSe_2$ monolayers. The larger of the two separations corresponds to the equilibrium distance between $CH_4$ and the surface for each surface. The $CH_4$ molecule was displaced along multiple paths, each divided into ten equal segments, to assess the direction-dependence of the sliding energy. Figure 5 presents the sliding energetics, showing that $CH_4$ encounters a significantly larger sliding barrier on the SnSe monolayer at equilibrium separation than $SnSe_2$, consistent across all directions examined. This result reinforces our experimental findings, demonstrating a higher resistance to sliding for SnSe than $SnSe_2$. Even when the $CH_2$ molecule is moved closer to the surface by 0.5 Å, the trend remains unchanged: $CH_4$ continues encountering a more substantial sliding barrier on SnSe than $SnSe_2$.

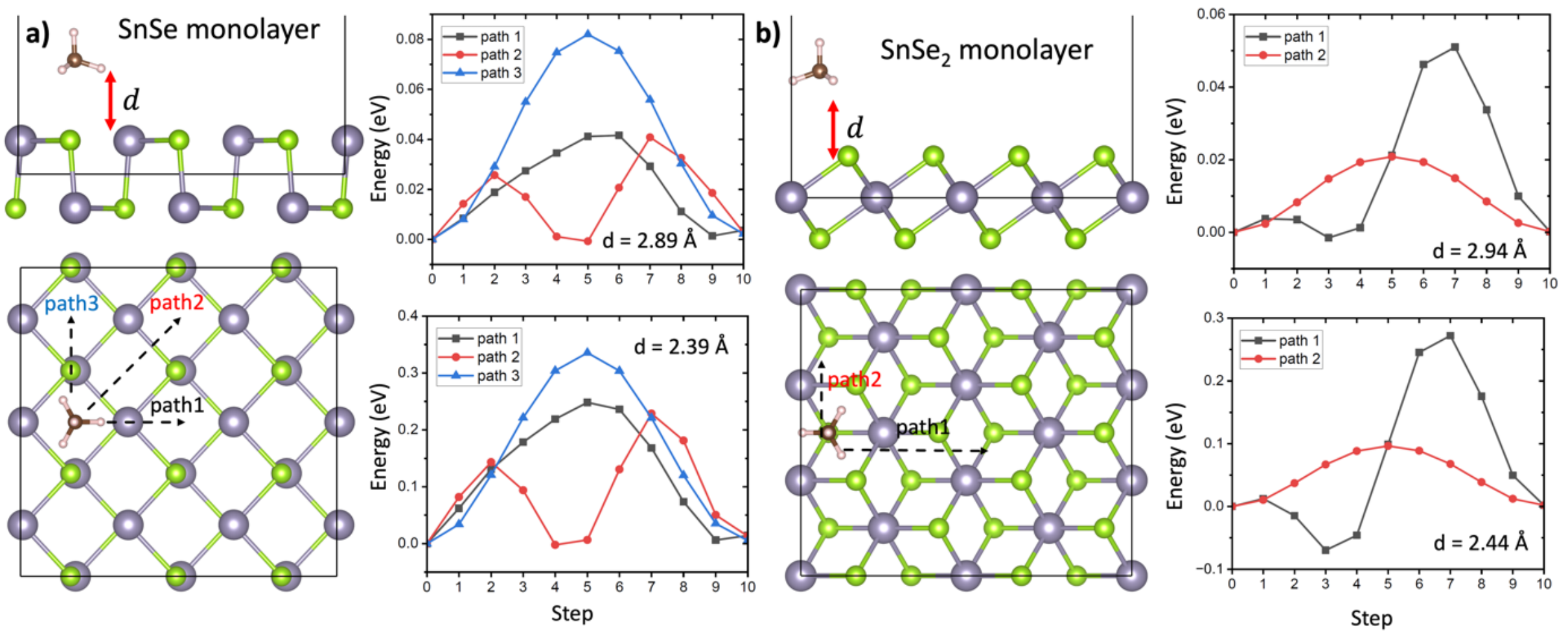

**Figure 5.** Sliding energy barriers of a $CH_4$ molecule over SnSe and $SnSe_2$ monolayers along different paths at two separation distances. **(a)** Results for the SnSe monolayer: the upper left shows a side view schematic of the $CH_4$ molecule on the SnSe monolayer at a distance d, and the bottom left displays a top view of three different paths traced by the $CH_4$ probe. The upper right shows the sliding energy barriers corresponding to the paths at a separation distance of 2.89 Å, while the bottom right shows the energy barriers at a separation of 2.39 Å. **(b)** Results for the $SnSe_2$ monolayer: the upper left depicts a side view of the $CH_4$ molecule on the $SnSe_2$ monolayer at a distance d, with the bottom left showing the top view of the three paths followed by the $CH_4$ probe. The upper right shows the sliding energy barriers at a separation of 2.94 Å, and the bottom right presents the energy barriers at a separation of 2.44 Å.

## Conclusion

We conducted LFM measurements and DFT calculations to investigate the frictional properties of SnSe and $SnSe_2$ nanoflakes. LFM results showed that SnSe and $SnSe_2$ flakes exhibited superior lubricative performance compared to graphene flakes of similar thickness, with SnSe flakes having nearly three times lower friction than graphene. Conversely, bulk SnSe showed higher friction than bulk graphite. DFT calculations further revealed that the interlayer binding energy of SnSe layers is greater than that of $SnSe_2$. To determine whether this higher binding energy correlates with increased friction, we compared the potential energy surfaces (PES) for bilayer SnSe and $SnSe_2$ under three conditions: (1) sliding of the top layer over the bottom layer at the equilibrium interlayer separation without relaxation, (2) sliding with the topmost and bottommost atoms fixed while allowing other atoms to relax, and (3) sliding with relaxation only in the out-of-plane (z-direction). Across all conditions, although PES patterns were similar for both bilayers, SnSe consistently showed significantly higher friction than $SnSe_2$, matching the calculated adhesion energies. Lastly, we simulated the sliding of a $CH_4$ molecule on monolayer SnSe and $SnSe_2$ surfaces in various directions. The results indicated greater energy dissipation on the SnSe surface in all directions, highlighting a strong link between friction and the differing atomic arrangements of the surfaces.

## Supplementary

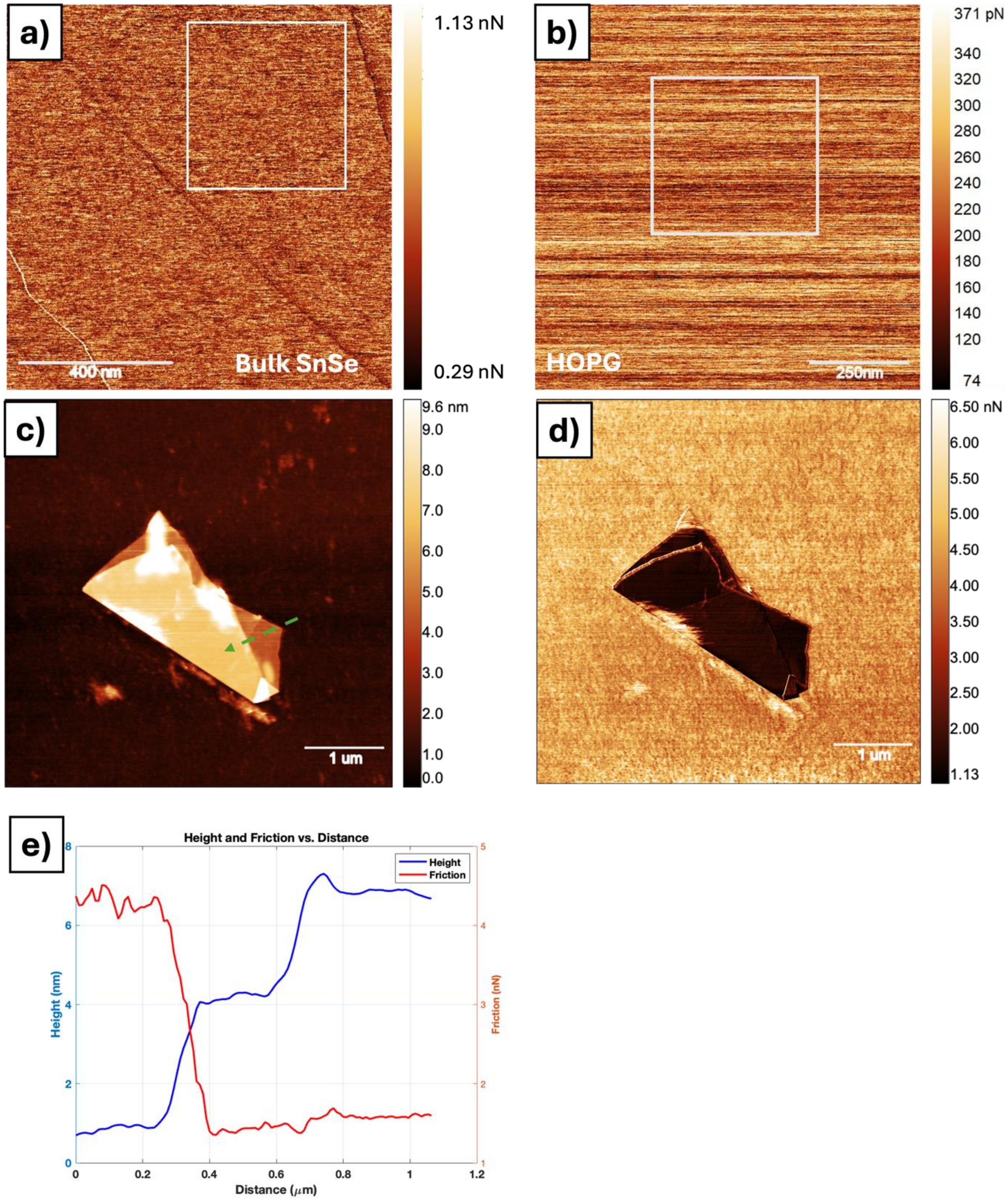

***Figure S1.*** Friction and topography analysis of Bulk SnSe, HOPG, and few-layer graphene. **(a)** Friction map of bulk SnSe, with a white rectangle indicating the area where the average friction force was measured. **(b)** Friction map of highly oriented pyrolytic graphite (HOPG), with a white rectangle highlighting the area used for average friction force calculations. **(c)** Topography map of few-layer graphene, height difference

contrast. **(d)** Corresponding friction map of the few-layer graphene. **(e)** The green indicates a comparison of height and friction profiles from the region dashed arrow in (c).